# How Can Journal Impact Factors be Normalized across Fields of Science?

# An Assessment in terms of Percentile Ranks and Fractional Counts

Loet Leydesdorff,[a] Ping Zhou,[b] & Lutz Bornmann [c]

**Abstract**

Using the CD-ROM version of the *Science Citation Index* 2010 ($N$ = 3,705 journals), we study the (combined) effects of (*i*) fractional counting on the impact factor (IF) and (*ii*) transformation of the skewed citation distributions into a distribution of 100 percentiles and six percentile rank classes (top-1%, top-5%, etc.). Do these approaches lead to field-normalized impact measures for journals? In addition to the two-year IF (IF2), we consider the five-year IF (IF5), the respective numerators of these IFs, and the number of Total Cites, counted both as integers and fractionally. These various indicators are tested against the hypothesis that the classification of journals into 11 broad fields by PatentBoard/National Science Foundation provides statistically significant between-field effects. Using fractional counting the between-field variance is reduced by 91.7% in the case of IF5, and by 79.2% in the case of IF2. However, the differences in citation counts are not significantly affected by fractional counting. These results accord with previous studies, but the longer citation window of a fractionally counted IF5 can lead to significant improvement in the normalization across fields.

[a] Amsterdam School of Communications Research (ASCoR), University of Amsterdam, Kloveniersburgwal 48, 1012 CX Amsterdam, The Netherlands; loet@leydesdorff.net; http://www.leydesdorff.net
[b] Institute of Scientific and Technical Information of China, 15th Fuxing Road, Haidian District, Beijing, China. Email: zhoup@istic.ac.cn
[c] Max Planck Society, Administrative Headquarters, Hofgartenstraße 8, 80539 Munich, Germany; bornmann@gv.mpg.de

**Introduction**

Although the journal impact factor (IF) is still widely used for journal evaluation, this measure has become increasingly debated to the extent that *Scientometrics* is preparing a special issue about flaws in both the construction and the use of IFs (Vanclay, 2012, in press). In this context, Leydesdorff (2012a, in press) suggests that two previously made proposals could be combined: (1) the use of fractional counting of citations in order to correct for differences among fields and journals in citation practices (Garfield, 1979; Moed, 2010) and (2) the use of non-parametric statistics because of the highly skewed distributions of citations at the level of both journals and fields (Seglen, 1992, 1997). In two previous studies (Leydesdorff & Bornmann, 2011a and b), we addressed these issues separately using IFs 2008 and 2009, respectively, for the comparisons, and based on different journal sets. In this study, we use the single set of 3,705 journals contained at the CD-ROM version of the *Science Citation Index* (SCI) 2010. Which combination of normalizations is optimal, and why?

**Theoretical relevance**

Garfield (1972; cf. Sher & Garfield, 1955) introduced the IF as a tool for journal evaluation with the natural and life sciences in mind. These sciences have a rapid turnover of citations and publications at the research front (Bensmann, 2007; Martyn & Gilchrist, 1968). The IF (IF2) focuses on the most recent two years, and numbers of citations within the current year are averaged with reference to the number of citable items published in the journal during the previous two years. Following the expansion of citation indices into the social sciences with the launch of the *Social Science Citation Index* (SSCI) in 1978, analysts have become increasingly aware that publication and citation practices differ among disciplines, and that a standard normalization such as IF2 might hide more than it reveals of the quality of the research in question (Price, 1970). In an early attempt to cluster the SCI, for example, Carpenter & Narin (1973) noted that multi-disciplinary journals such as *Science* and *Nature* are very differently positioned in the journal hierarchy than disciplinary or specialist journals. How should one account for these differences among specialties and types of journals (Leydesdorff, 2008)?

In response to the critique that a two-year citation window might be too short for disciplines with slower turn-overs, Thomson-Reuters (TR) extended the IF with a five-year variant (IF5). A host of journal indicators also became available with the emergence of the new network sciences in the 2000s. Using factor analysis on a set of journal indicators (for 8000+ journals in 2008), however, Leydesdorff (2009) showed that the IFs among the different network indicators provide an independent dimension *because of the division* by the number of publications. One can consider the IF as an average number of citations of a journal during the last two (or five) years, whereas the other indicators do not assume this parametric normalization.

Using averages of highly skewed distributions, however, may cause problems and misinterpretations. Seglen (1992, 1997) was the first to argue that one should avoid taking the mean (or any other central tendency statistics) over a highly skewed distribution such as the citation distributions of journals. Bornmann & Mutz (2011) intervened in the discussion (in the *Journal of Informetrics*; cf. Gingras & Larivière, 2011) about dividing averages or averaging rates by proposing the metrics of the six percentile rank classes in use for the *Science &*

*Engineering Indicators*: top-1%, top-5%, top-10%, top-25%, top-50%, and bottom-50% as an alternative to using averages (National Science Board, 2012).

Leydesdorff *et al*. (2011) elaborated the non-parametric statistics for percentile rank classes, and Leydesdorff & Bornmann (2011b) then applied a newly defined "Integrated Impact Indicator" (*I3*) to two groups of journals: the set of 65 journals classified in the Web of Science (WoS) as Library & Information Science, and the 48 "multidisciplinary" journals, including journals such as *Science, Nature,* and *PNAS*. Such non-parametric perspectives can lead to statistically significantly different rank-orderings when compared with parametric ones. However, these studies (including Leydesdorff, 2012b, in press) were based on using specific WoS Subject Categories (WC) for the delineation of groups of journals in terms of fields of science.

WCs are attributed at the journal level, and the sets thus delineated may overlap in terms of the journals involved. Rafols & Leydesdorff (2009) have shown that the attributions of WCs to journals match poorly with algorithmic journal classifications based on aggregated journal-journal citation relations using the Journal Citation Reports (JCR) as the data source. Pudovkin & Garfield (2002, p. 1113) expressed earlier doubts about the analytical quality of these attributions: "Journals are assigned to categories by subjective, heuristic methods. In many fields these categories are sufficient but in many areas of research these 'classifications' are crude and do not permit the user to quickly learn which journals are most closely related."

Boyack (personal communication, 14 September 2008) estimated that approximately 50% of the attributions of WCs to journals were misplaced, but Rafols *et al*. (2010) argued that these misplacements are random and rarely more than a single category away in the map of science. Therefore, a global map based on these attributions might still be reliable at the aggregated level. However, Boyack & Klavans (2011) have argued that journals themselves should not be considered as mono-disciplinary units even if the attribution of WCs were unambiguous. Attribution of disciplinary identities *at the level of papers* would thus be preferable (Bornmann *et al.*, 2008; Lundberg *et al.*, 2006).

The systematic and significant differences between disciplines in the citation impacts of their journals find their origin in differences in citation behavior among fields (Bornmann & Daniel, 2008; Garfield, 1979). In some fields (e.g., mathematics) citation lists are short, while in others (e.g., the biomedical sciences) long reference lists prevail. The number of references also varies significantly among document types. Reviews have typically long reference lists and longer-term citation value (Price, 1965),[1] whereas Letters to the Editor involve rapid communication. Citation turn-over thus varies both among fields and document types. There are also changes across the fields over time (Althouse *et al.*, 2009).

Following up on a suggestion of Moed (2010), Leydesdorff & Opthof (2010a and b) proposed to explore the normalization of citations in terms of the length of reference lists, that is, using 1/number of references in the *citing* papers as a method to correct for the differences in citation behavior among fields of science. Note that this normalization is paper-based: one defines the relevant reference set ("the field") as the set of citing papers without an a priori classification.

---

[1] "In the *JCR* system any article containing more than 100 references is coded as a review. Articles in 'review' sections of research or clinical journals are also coded as reviews, as are articles whose titles contain the word 'review' or 'overview.'" At http://thomsonreuters.com/products_services/science/free/essays/impact_factor/ (retrieved April 8, 2012).

Such fractional counting of the citations had been proposed by Small (1985) for the generation of co-citation maps, and was placed back on the table by Zitt & Small (2008) when these authors proposed the Audience Factor among citing journals as an alternative to the Impact Factor of the journals cited. Moed (2010) identified fractional counting as a possible operationalization of Garfield's (1979) notion of "citation potentials." Fields with (on average) long reference lists provide authors more chances of being cited than fields with short reference lists (such as mathematics).

Similarly, there can be a dynamic effect since some fields have faster or slower research fronts than others in terms of publication delays and citation windows. Could one also correct for these dynamic field differences in citation behavior by counting each citation a $1/(n$ of references)? Glänzel *et al.* (2011, at p. 412) suggested that the two effects have to be studied using two completely different statistics: these authors proposed to measure the static effect as the conditional mean of the fractionally counted citations (given a citation window), and the dynamic one as the relative frequency of the uncited papers. Using a three-year citation window (2006-2008), the conditional means of fractionally counted citation rates was then found to be almost subject-independent using the twelve major science fields distinguished in the Leuven classification scheme (Glänzel & Schubert, 2003). Using data of the CD-ROM version of the SCI 2008—*n* of journals = 3,853—and the 13 main fields defined by PatentBoard/NSF for the purpose of organizing the statistics in the *Science and Engineering Indicators* of the U.S. National Science Board, Leydesdorff & Bornmann (2011a) similarly showed that a quasi-IF2 on the basis of fractionally counted citations reduces the between-field variance by 81.3%, while the remaining differences among the fields were no longer statistically significant.[2]

Thus, the static effect was confirmed, but the conclusions of both studies remained inconclusive about the dynamic effect (perhaps because of the use of different statistics). If one uses a two-year citation window as in the case of IF2, many biomedical journals will have more than 50% of the references within this window, while in mathematics (or the social sciences) the figure will be less than 50%. The distribution of 1/NRef for all years included in NRef is (potentially significantly) different from the 1/NRef when only the last two or five years are taken into account. Accordingly, one can normalize using only the last two (or five) years as a restricted reference sets, or using the total number of references (NRef) as provided in WoS.[3]

Including all previous years, Leydesdorff & Bornmann (2011b) found that a fractionally counted c/p ratio (2008) also reduced the between-field variance to the extent that the differences were no longer statistically significant, but less so than in the case of using a fractionally counted IF2. As noted, the reduction was 81.3% in the latter case, but only 41.7% of the between-field variance was reduced using fractionally counted c/p-ratios. Thus, the inclusion of the dynamic effect would, according to these authors, not improve the normalization among fields using fractional counting.

---

[2] An Excel sheet with these quasi-IF2s 2008 is available at http://www.leydesdorff.net/weighted_if/weighted_if.xls.
[3] De Andrés (2011) additionally suggested using the average value of NRef for each WC, respectively. In our opinion, this approach combines the a priori (source-based) and a posteriori (indexer-based) normalizations that were analytically distinguished by Glänzel et al. (2011).

Perhaps the normalization is also affected by taking the mean over a skewed distribution. In this study, we therefore also consider the numerators of the IFs, both fractionally and integer counted, and using two or five-year time windows. Bensmann *et al.* (1996) argued that "total times cited" (TC) could be validated as an indicator of journal prestige more than IFs. TCs can be counted fractionally, and one can limit the citation window to two or five years in order to generate the corresponding numerators of (quasi-)IFs. Leydesdorff (2012a, in press) concludes that it should then be possible to combine the two ideas: fractional counting and non-parametric statistics at the level of the complete journal set (cf. Beirlant *et al.*, 2007).

Instead of returning to the data of 2008 used as the latest available to Leydesdorff & Bornmann (2011a), we decided to replicate the analysis from scratch using similar data for 2010. Whereas we previously were constrained in our handling of the approximately 30 million references in relational database management, in the meantime the limits on file sizes are no longer in place under Windows 7 on a 64-bits operating system. Thus, in addition to the possibility to compare integer with fractionally counted quasi-IFs, the design allows us to compare different time horizons, and the percentile rank classes of all possible IFs and their numerators, that is, citation counts. The various indicators are used as dependent variables for testing using multi-level regression models in Stata.

**Methods and materials**

The CD-ROM version of the SCI 2010 contains 973,182 documents organized under 3,725 journal names. These documents contain 27,892,190 cited references. This data was downloaded and reorganized in a relational database management system so that dedicated queries could be written for organizing the data in the various desired output formats.

Of the 3,725 journal names, 24 did not correspond with the 8,005 journal names in the JCR 2010 of the SCI-Expanded edition of the same index, available online at the Web of Knowledge. Three mismatches find their origin in incidental misspellings. Furthermore, the *Journal of Geophysical Research* and *Tissue Engineering* are split into seven and three parts in the CD-ROM version, respectively, but not at WoS. We first counted these journal parts separately, then merged them thereafter for comparisons with the ISI-IFs. Seven journal names are not included in WoS, but are counted as such on the CD-ROM version (e.g., *Abstracts of Papers of the American Chemical Society*). In sum, our comparisons are based on 3,705 journals.

The approximately thirty million cited references can be parsed into subfield information, such as the venue abbreviation and the publication year of each reference. A subset of the 1,152,760 venue abbreviations could be matched unambiguously with the 3,705 full journal names in the bibliographic headings of the citing records. Thereafter, routines can be used for counting and fractionating the references for the last two years (2008 and 2009), the last five years (2005-2009), or all years.

The CD-ROM version is based on data entered into the system between January 1 and December 31 of 2010. This includes 8.7% (84,334) records with a publication year 2009, and smaller numbers for other years. The number of publications with publication year 2010 is 864,697 or 88.9% of the 973,181 records retrieved. The quasi-IFs which can thus be derived from this data, cannot be compared directly with the ISI-IFs in the JCR because the latter are based on 2010 data as defined at a cutoff in March of the following year by the staff of TR (McVeigh, *personal communication*, April 7, 2010) and numerators of the ISI-IFs are counted in the full set of 10,196 journals included in the JCRs of the SCI and SSCI. Note that JCR includes a larger set of journals in the SCI-Expanded version at WoS.

|  | All years | 2005 to 2009 | 2008 and 2009 |
|---|---|---|---|
| a. JCR version | 30,731,845 | 12,084,789 | 4,834,067 |
| b. CD-ROM version | 27,852,549 | 10,780,462 | 4,419,598 |
| c. Reference within the $N = 3,705$ subset | 19,989,208 (65.0%) | 8,173,899 (67.6%) | 3,352,108 (69.3%) |
| d. As c., but fractionally | 548,066 | 552,638[4] | 503,294 |

**Table 1**: Citation counts for different journal sets and time spans.

Among the 27,892,190 cited references, 39,641 (0.14%) do not contain a valid (numeric) date. Another 0.6% contains a date smaller than 1900. (We had programmed the routine to test for whether the date contained a valid first two digits "19" or "20".) This left us with 27,852,549 (99.86%) valid references (Table 1, row b). Of these references, 4,419,598 (15.87%) contain 2008 and 2009 as publication years (the two preceding years) and 10,780,462 (38.71%) refer to the five preceding years (2005-2009).

References were summed both fractionally (1/NRef; row d in Table 1) and as integers (row c) for the three periods of 2008+2009 (two years), 2005-2009 (five years), and all years. Row c of Table 1 informs us that the subset of 3,705 journals (36.4%) accounts for 65.0% of the citations. These citations are concentrated in the last two years (69.3%) even more than in the last five years (67.6%). This confirms the claim of PatentBoard, the NSF, and TR that these 3,705 journals can be considered as a core group within SCI-Expanded.

---

[4] When fractionated with reference to the total number of references (that is, including all publication years in the references), FC2+ = 98,623 and FC5+ = 228,826.

Fractionally counted, the numbers are normalized (row d in Table 1). For all years, for example, this means that 19,989,208 integer counted citations are equal to 548,066 fractionally counted citations, and thus the average number of references (NRef)/paper is 19,989,208/548,066 = 36.47. The equivalent numbers for the last five or two years are 14.79 and 6.66, respectively. Of the references, 40.9% are attributed to the last five years, and 16.8% to the last two years.

From this data, quasi-IF numerators can be constructed; IF denominators were harvested from the JCR 2010. Both the numerators and the quasi-IFs can thereafter also be transformed into percentile distributions.

| | Variable | 100 percentile ranks (*PR100*) | 6 percentile ranks (*PR6*) |
|---|---|---|---|
| 1. | ISI-IF2 | | |
| 2. | ISI-IF5 | | |
| 3. | IF2-IC | | |
| 4. | IF5-IC | | |
| 5. | IF2-FC | | |
| 6. | IF5-FC | | |
| 7. | IF2-FC+ | | |
| 8. | IF5-FC+ | | |
| 9. | fc/p | + | + |
| 10. | ISI-TC | + | + |
| 11. | TC-IC | + | + |
| 12. | TC-IC2 | + | + |
| 13. | TC-IC5 | + | + |
| 14. | TC-FC | + | + |
| 15. | TC-FC2 | + | + |
| 16. | TC-FC5 | + | + |
| 17. | TC-FC2+ | + | + |
| 18. | TC-FC5+ | + | + |
| 19. | IF2-Num | + | + |
| 20. | IF2-Denom | + | + |
| 21. | IF5-Num | + | + |
| 22. | IF5-Denom | + | + |
| 23. | Items2010 | | |

**Table 2**: Variables and Percentile Ranks included in the analysis

Table 2 summarizes the variables that will be used in the testing. The 23 variable names distinguished in this study are listed (as abbreviations) in the second column. Variables 1 and 2 are the ISI Impact Factors for two and five years (ISI-IF2 and ISI-IF5), respectively, based on the 10,196 journals included in WoS. IF2-IC and IF5-IC denote the quasi-IFs corresponding to these, but based on the 3,705 journals in our sample. IF2-FC and IF5-FC provide the same two values, but based on fractional counting of the citations. These values are based on fractionation by the numbers of references within the two (or five) preceding years. However, one can also consider all references in the citing paper as the value of this denominator. The latter normalization provides IF2-FC+ and IF5-FC+, where the + indicates that the larger set of all references is considered relevant. All quasi-IFs are based on using the IF2-Denominator and the IF5-Denominator values provided in JCR for the division. Numerators of the (quasi-)IFs are provided as variables in rows 10-19 and 21 using a notation analogous to the one above for IFs, but using "TC" instead of "IF".

One additional variable is defined in row 9: fractional citations/publications. This variable is the result of dividing the total of fractional citation counts (variable 14) by the total number of citable items in 2010 (variable 23). Thus, this indicator can be considered as a fractional c/p ratio for 2010. We used this variable before (Leydesdorff & Bornmann, 2010a) to test whether, in addition to the fractionally counted IF, a dynamic term should be distinguished for the normalization across different disciplines (cf. Glänzel *et al.*, 2011). The IF2 focuses only on the last two years, the IF5 on only the last five years, but in some fields older citations are more important than more recent ones (Ludo Waltman, *personal communication*, 23 June 2010).[5] However, the c/p ratio includes all the previous years; by fractionally counting this ratio, both the static and the dynamic effects can be captured as a single indicator.

Percentile rank classes will be reported for times cited (that is, numerators) either fractionally or integer-based. The quantile values (underlying the percentile rank classes) are determined in this study by using the counting rule that the number of items with lower citation rates than the item under study determines the percentile value (Leydesdorff & Bornmann, in press). Tied citation numbers are thus accorded the same values. (See for other approaches to tied ranks, for example: Pudovkin & Garfield, 2009; Rousseau, 2012; Schreiber, 2012a). The purpose is to transform the skewed citation distributions among journals into a percentile distribution, and to study the effects of this transformation on the comparison among journals. In addition to the percentiles (*PR100*), *PR6* values are derived by aggregating the quantile values using the six classes of the National Science Board (2012) specified above, so that quantiles ≥ 99 are counted into class 6, quantiles ≥ 95 into class 5, etc. (Bornmann & Mutz, 2011). This transformation is non-linear and therefore the *PR6* values will be tested independently from the *PR100* values.

---

[5] Waltman & Van Eck (2010) suggested an additional normalization based on the *average* number of references in the citing journal.

In addition to studying the substantive effects of these various options on the rankings of the journals and their numerical effects on the correlations among these ranks, we test the effects of the grouping into the 13 main-field categories that have been used for the construction of the *Science & Engineering Indicators* by the NSF and PatentBoard (formerly ipIQ or CHI Inc.). We chose this classification because it is reflexively shaped and updated regularly on a journal-by-journal basis without automatic processing (cf. Rafols & Leydesdorff, 2009). Furthermore, journals are uniquely attributed to a broad field.

The effect of the grouping can be tested in terms of the between-group variance in relation to the within-group variances in the 13 classes. Using ANOVA, for example, one can study $\eta^2$ as a measure of this between-group variance ($\eta^2 = SS_{effect} / SS_{total}$; Cohen, 1977, pp. 280 ff.), the Interclass Correlation Coefficients (ICC) or analyze variance components using "mixed models" in SPSS. However, these routines assume normality of the distribution in the dependent variable. Using tests for normality of the distribution (Kolmogorov-Smirnov and Shapiro-Wilk in SPSS), this assumption was rejected for all the variables under study. Therefore, we decided to use other statistical procedures.

Depending on the scale of the dependent variable, we tested three different two-level regression models:

1. IF2, IF5 and Times Cited metrics are based directly on citation counts for the papers published in the journal and therefore follow the typical skewed distribution of citation data. Thus, they can be treated as citation counts. Since a Poisson distribution is often appropriate in the case of count data (Cameron & Trivedi, 1998), we calculated a two-level random-intercept Poisson model for these journal metrics. To handle overdispersion at level 1 (measured by large differences between the mean and the variance of the IFs and Total Cites) in a two-level random-intercept Poisson model, Rabe-Hesketh and Skrondal (2008, p. 395) recommend the use of the sandwich estimator for standard errors.
2. For journal metrics based on *PR100* we use simple variance-components models with the robust option of Stata (White, 1980). We use this robust option, since the histograms of these metrics show distributions that are similar to a normal distribution, but the tests for normality show statistically significant deviations from the normal distribution (see above). Thus, we calculate robust standard errors based on the sandwich estimator, as these do not depend on the model being specified correctly.
3. Since the journal metrics based on *PR6* are categorical variables with ordered categories, we have calculated random-intercept proportional odds models (Rabe-Hesketh & Skrondal, 2008, chapter 7.6).

Two fields in the classification ("Humanities" with n = 2 and "Professional Fields" with n = 8) had such low numbers of journals that they had to be excluded from the computation of the regression models. We estimate two-level models, where the IF (and TC) metrics are (3,695) level-1 units and the (11) fields are level-2 clusters. The random intercept is referred to as the level-2 residual with level-2 (between-field) variance. The higher the coefficient of the variance component, the larger is the difference between the fields according to the journals' citation impacts. The log-likelihood ratio is used to test whether two models (one with and one without random intercept) differ. If this test shows a statistically significant result, this difference signals a systematic field-specific difference between the journal metrics across the 11 broad fields. Statistical significance will be tested for these models at the level of $p < 0.001$ because of the large number of journals in the study.

In summary, 3,695 journals as cases are clustered into 11 fields. The objective of all statistical procedures is to investigate whether the differences in citation behavior among these 11 fields remain significant despite the corrections on the distributions by fractional counting of the citations and/or the transformation into percentile rank classes.

**Differences from the previous study**

This study differs from the previous one (Leydesdorff & Bornmann, 2011a) mainly because our objective here is to distinguish analytically among the various effects, and to include a longer time horizon. In the previous study, we wished to show the effect of fractional counting using IF2 (as the standard impact indicator of journals). As noted, the availability of improved hard and software has made further refinements possible, and advancements during last year in the use of non-parametric statistics (Leydesdorff *et al*., 2011; Leydesdorff & Bornmann, 2011b) made it worthwhile to repeat the effort of data collection. The results of Glänzel *et al*. (2011) encourage this approach; unlike these authors, however, we prefer to test percentiles instead of average values in the case of skewed distributions (cf. Glänzel, 2010).

The data collection is similar to the previous case, but using SCI 2010 instead of SCI 2008. In Leydesdorff & Bornmann (2011a) we used only two-year IFs (Model 3) and c/p ratios (Model 4) as fractionally counted alternatives to integer counted IFs. Unlike ISI-IF2s (Model 1) and integer-counted quasi-IF2 (Model 2), Models 3 and 4 were no longer statistically significant in that study using the variance component analysis described above. The fractionally counted IF2s reduced the between-group variance by 81.3% and the remaining between-group variance was no longer statistically significant. The fractionally counted c/p ratios reduced the between-group variance by only 41.7%.

In the previous study, we used the abbreviations in the cited references for all (at the time, 6,598) journals included in the Science Edition of JCR 2008. A match could be found in the case of 5,794 (87.8%) of these journals. However, the data was harvested from the CD-ROM version containing 3,853 citing journals in 2008. For various reasons, many of these additional journals had to be excluded in later stages of the analysis. In this study, therefore, we decided to focus exclusively on the set of 3,705 journals contained in the CD-ROM version of the SCI, both cited and citing. This restriction may reduce error because of mismatches and incomplete data. Note that the number of journals included in the CD-ROM version has shrunk from 2008 to 2010, whereas the JCR has been expanded. As in the previous study, we use the 13 categories of PatentBoard/NSF which were kindly provided to us by these two organizations; but in the case of the multi-level regression models using Stata, two categories ("Humanities" with n = 2 and "Professional fields" with n = 8) were not considered, so that N = (3705 – 10) = 3695.[6]

In addition to extending the study to IF5, we distinguish in this study between two ways of fractionating the citation counts. As in the previous study, IF2-FC and analogously IF5-FC are first considered by using as the denominator all references to the two (five) previous years only, for each respective article. Note that it is not possible to control for document types among the cited references, but this is also the case for the construction of the numerators of the ISI-IFs at TR (Moed & Van Leeuwen, 1996).

Secondly, it is possible to construct a fractionally counted citation rate based on using the *total* number of references in each citing paper as the denominator, irrespective of the publication year of the cited reference. (This is the number which is provided as "NRef" in the bibliographic information of WoS.) We indicate these two (lower!) values as IF2-FC+ and IF5-FC+, respectively. In the previous study we used also the fractionally counted c/p ratio. Note that the denominator of the c/p ratio is provided by the number of citable items in the current year (2010), whereas in the IFs and quasi-IFs, the denominator corresponds to the JCR denominator value (based on summing citable items for the two or five previous years).

---

[6] In 2008, 41 journals were classified as "Professional Fields" and 2 as "Humanities". We used 13 categories at the time, but removed these two categories in this analysis because of their small sizes.

**Results**

Let us turn first to the IFs and focus thereafter on the total citations and the numbers used as numerators in the various IFs and quasi-IFs.

*a. Fractional counting of the impact factor*

|          | ISI-IF2 | ISI-IF5 | IF2-IC | IF5-IC | IF2-FC | IF5-FC | IF2-FC+ | IF5-FC+ | FC/P |
|----------|---------|---------|--------|--------|--------|--------|---------|---------|------|
| ISI-IF2  |         | .972    | .915   | .899   | .857   | .815   | .850    | .868    | .630 |
| ISI-IF5  | .977    |         | .877   | .901   | .826   | .824   | .807    | .854    | .646 |
| IF2-IC   | .960    | .960    |        | .978   | .946   | .898   | .912    | .915    | .678 |
| IF5-IC   | .936    | .968    | .984   |        | .928   | .925   | .877    | .926    | .705 |
| IF2-FC   | .933    | .922    | .959   | .934   |        | .958   | .927    | .932    | .725 |
| IF5-FC   | .906    | .932    | .947   | .958   | .973   |        | .898    | .957    | .765 |
| IF2-FC+  | .930    | .917    | .946   | .917   | .965   | .949   |         | .964    | .698 |
| IF5-FC+  | .923    | .944    | .955   | .959   | .958   | .982   | .977    |         | .738 |
| FC/P     | .749    | .783    | .763   | .786   | .788   | .827   | .778    | .806    |      |

**Table 3**: Spearman rank-order correlations (upper triangle) and Pearson correlations (lower triangle) among IFs and quasi-IFs; $N = 3,705$; all correlations are significant at $p < 0.001$.

Table 3 shows the Spearman rank-order correlations among the various IFs and quasi-IFs (see Table 2) in the upper triangle and the Pearson correlation coefficients in the lower triangle. All correlation coefficients are statistically significant at $p < 0.001$ because of the large number of journals involved ($N = 3,705$). We focus in the following on the rank-order correlations because (as noted) all distributions were statistically significantly different from normal.

With the exception of fc/p, the metrics of the corresponding quasi-IFs are highly correlated ($\rho$ is in the order of 0.8 - 1.0). The correlations between the respective quasi-IFs using a two versus five-year window are marginally higher than between the corresponding integer or fractionally counted quasi-IFs. For example, Spearman's $\rho$ between IF2-FC and IF5-FC is 0.958, whereas the IF5-IC and IF5-FC correlate with $\rho = 0.925$. In other words, the change between integer and fractional counting is always marginally larger than between using the two different citation windows.

By enlarging the citation window to all previous years (in the case of fc/p) a somewhat different indicator is generated: Spearman's rho with the other indicators is in the order of 0.7. As could be expected, including all years in the fractionation (as in IF2-FC+ and IF5-FC+) or focusing only on the last two or five years shows a greater effect for two years than for five years. These latter effects are marginal at the level of the distributions. Still, there may be important effects for the ranking of individual journals.

|    | IF5-IC (a)          | IF5-FC (b)          | IF5-FC5+ (c)        | FC/P (d)            |
|----|---------------------|---------------------|---------------------|---------------------|
| 1  | Rev Mod Phys        | Rev Mod Phys        | Ca-Cancer J Clin    | New Engl J Med      |
| 2  | Ca-Cancer J Clin    | Ca-Cancer J Clin    | Rev Mod Phys        | Prog Solid State Ch |
| 3  | Annu Rev Immunol    | New Engl J Med      | New Engl J Med      | Adv Phys            |
| 4  | New Engl J Med      | Chem Rev            | Nat Mater           | Annu Rev Immunol    |
| 5  | Chem Rev            | Nat Mater           | Nat Photonics       | Lancet              |
| 6  | Nat Rev Mol Cell Bio| Annu Rev Immunol    | Annu Rev Immunol    | Phys Rep            |
| 7  | Nat Mater           | Physiol Rev         | Nat Nanotechnol     | Annu Rev Biochem    |
| 8  | Physiol Rev         | Nat Photonics       | Lancet              | Nature              |
| 9  | Annu Rev Biochem    | Prog Mater Sci      | Chem Rev            | Rev Mod Phys        |
| 10 | Nature              | Nat Rev Mol Cell Bio| Jama-J Am Med Assoc | Jama-J Am Med Assoc |
| 11 | Nat Rev Cancer      | Lancet              | Nat Rev Mol Cell Bio| Science             |
| 12 | Nat Nanotechnol     | Phys Rep            | Nature              | Chem Rev            |
| 13 | Cell                | Annu Rev Biochem    | Nat Rev Cancer      | Physiol Rev         |
| 14 | Nat Rev Immunol     | Nature              | Physiol Rev         | Adv Catal           |
| 15 | Nat Rev Neurosci    | Jama-J Am Med Assoc | Prog Mater Sci      | Ca-Cancer J Clin    |
| 16 | Science             | Nat Nanotechnol     | Science             | Prog Mater Sci      |
| 17 | Annu Rev Neurosci   | Surf Sci Rep        | Annu Rev Biochem    | Surf Sci Rep        |
| 18 | Annu Rev Astron Astr| Science             | Nat Genet           | Cell                |
| 19 | Nat Genet           | Adv Phys            | Nat Rev Drug Discov | Annu Rev Neurosci   |
| 20 | Nat Photonics       | Nat Rev Cancer      | Mat Sci Eng R       | Endocr Rev          |

**Table 4**: The top-20 journals in terms of IF5-IC, IF5-FC, IF-FC5+, and FC/P compared.

Table 4 shows the top-20 journals when ranked according to the integer counted impact factor for five years (IF5-IC) in column a, the fractionally counted equivalents in columns (b) and (c) but using two different modes of fractionation, and the fractionally counted c/p ratio in column (d). Although the lists show considerable overlap (and the correlations at the level of the distributions were high), the different rankings can matter considerably for individual journals. For example, *CA-A Cancer Journal for Clinicians* with ISI-IF5 of 70.245 is second behind the *Reviews of Modern Physics* (ISI-IF5 = 48.621) when the domain is restricted to these 3,705 journals. However, this journal ranks only at 15[th] position on the fc/p indicator.

The large journals with a multidisciplinary scope are better positioned if the citation window is not restricted. The *Lancet*, for example, ranks 6[th] in column d, but only 8[th] and 11[th] in columns c and b, respectively. Using the quasi-IF5, integer counted, for the CD-ROM version (column a), the *Lancet* is no longer included among the top-20 (in column a), but ranks at the 29[th] position.[7] In summary and as expected, a choice among the different options remains debatable. Let us therefore turn to the variance components models.

---

[7] Using ISI-IF5, the *Lancet* ranks at the 12[th] position in this set of 3,705 journals. The *Lancet*, in other words, earns relatively more credit from the journals in WoS that are not included in the CD-ROM set.

*b. The variance components models for the (quasi-)impact factors*

| Model | Variable | Level 2 variance component (with S.E.) |
|---|---|---|
| M1 | ISI-IF2 | .12 (.03)* |
| M2 | ISI-IF5 | .13 (.05)* |
| M3 | IF2-IC | .24 (.09)* |
| M4 | IF5-IC | .20 (.07)* |
| M5 | IF2-FC | .05 (.02)* |
| M6 | IF5-FC | .02 (.01) |
| M7 | IF2-FC+ | .07 (.03) |
| M8 | IF5-FC+ | .04 (.02) |
| M9 | fc/p | .06 (.03)* |

\* *p*<.001

**Table 5**: Variance component models for the IFs and quasi-IFs; $N = 3,695$; 11 broad fields. (Because of the large number of models involved, only the level-2 variance components are presented.)

Table 5 shows that the fractionally counted quasi-IFs (models M6 to M8) reduce the between-group variance to the extent that the differences among the 11 classes are no longer statistically significant. However, Model M5 which corresponds with the best model in the previous study (Model 3 in that study) fails to be statistically significant in this study. One can think of the more sharply delineated domain (see above) and further refinements in the classifications by PatentBoard/NSF as reasons for this different result. In summary, the five-year IF5-FC seems to be better suited for the purpose than the two-year IF2-FC. IF5-FC+ also reduces the between-group variance more than IF2-FC+.

The reading of the numeric results is as follows: In comparison with the integer-counted model M3 (IF2-IC), the level-2 variance component is reduced with M6 to $(0.24 - 0.02)/(0.24) = 0.22/0.24$ or 91.7%. This 91.7% outperforms the reduction of 81.3% that Leydesdorff &

Bornmann (2011a) found for the equivalent of M5 in the previous study. (M5 reduced between-group variance to 79.2% in this study.) However, M7 and M8—that is, including the references of previous publication years in the reference sets—do not improve on M6. The results of M9 confirm the finding that inclusion of older references (more than five years in the past) introduces dynamic between-field differences. In short: IF5-FC provides the best fit given our research question.

### c. Fractionally and integer-counted citations

|         | ISI-TC | TC-IC | TC-FC | TC-IC2 | TC-FC2 | TC-IC5 | TC-FC5 | TC-FC2+ | TC-FC5+ |
|---------|--------|-------|-------|--------|--------|--------|--------|---------|---------|
| ISI-TC  |        | .954  | .946  | .868   | .873   | .897   | .895   | .853    | .886    |
| TC-IC   | .984   |       | .976  | .921   | .917   | .950   | .937   | .889    | .924    |
| TC-FC   | .967   | .979  |       | .904   | .930   | .931   | .957   | .920    | .950    |
| TC-IC2  | .932   | .943  | .938  |        | .978   | .986   | .956   | .970    | .965    |
| TC-FC2  | .914   | .926  | .953  | .977   |        | .969   | .983   | .981    | .978    |
| TC-IC5  | .954   | .971  | .961  | .987   | .968   |        | .973   | .956    | .976    |
| TC-FC5  | .928   | .945  | .975  | .964   | .990   | .976   |        | .967    | .990    |
| TC-FC2+ | .903   | .908  | .951  | .968   | .987   | .956   | .980   |         | .985    |
| TC-FC5+ | .924   | .935  | .971  | .964   | .983   | .973   | .994   | .989    |         |

**Table 6**: Spearman rank-order correlations (upper triangle) and Pearson correlations (lower triangle) among Times Cited, fractionally and integer-based; $N = 3{,}705$; all correlations are significant at $p < 0.001$.

Table 6 provides the correlations between the citation counts in a format similar to Table 3 above. Note that these distributions and hence their correlations are different from those in Table 3 because each numerator of an impact factor is divided by the denominator of the respective journal; both numerators and denominators vary among journals. The correlations are again so high and the differences so marginal that we abstain from further interpretation.

| Model | Variable | Value | PR100 | PR6 |
|---|---|---|---|---|
| (a) | (b) | (c) | (d) | (e) |
| N1: | ISI-TC | .04(.01)* | 84.3 (40.7)* | .29 (.15)* |
| N2: | TC-IC | .14(.02)* | 130.1 (59.5)* | .48 (.24)* |
| N3: | TC-FC | .07(.02)* | 114.9 (71.5)* | .39 (.20)* |
| N4: | TC-IC2 | .13(.02)* | 165.9 (58.8)* | .77 (.36)* |
| N5: | TC-FC2 | .07(.02)* | 55.4 (6.4)* | .54 (.27)* |
| N6: | TC-IC5 | .15(.02)* | 152.5 (59.9)* | .72 (.33)* |
| N7: | TC-FC5 | .09(.01)* | 118.8 (62.6)* | .50 (.25)* |
| N8: | TC-FC2+ | .24(.03)* | 146.2 (56.1)* | .61 (.29)* |
| N9: | TC-FC5+ | .22(.02)* | 135.1 (61.4)* | .54 (.26)* |
| N10: | FC/P | .06(.03)* | 67.1 (47.5)* | .19 (.11)* |

* $p<.001$

**Table 7**: Level-2 variance components for total cites and percentiles; $N = 3,695$; 11 broad fields.

Table 7 shows that the level-2 variance components remain statistically significant in all these models. Thus, these variables are always field-specific in terms of the 11 broad fields distinguished by PatentBoard/NSF despite fractional counting. However, there is a reduction of differences using, for example, fractional counting with a two-year citation window: model N5 compared to model N2 reduces the level-2 variance component to (0.14 - 0.07)/0.14 or 50.0%. However, this reduction is not sufficient to make the field differences statistically no longer significant.

Note the relatively low value of the second level coefficient (.04) for the "total cites" in the ISI set (ISI-TC). Unlike all other variables, this value is based on all citations in the JCR domain (10,196 journals). In other words, in this journal set three times larger, the field differences among total cites of journals in the 11 classes, are reduced when compared to the more specifically selected journal set of SCI that is used for *Science & Engineering Indicators* and for this study. The smaller group of journals in the SCI is more field-specifically delineated in the citation patterns than the larger set. The larger set of SCI-Expanded can be expected to contain more interdisciplinary journals. Upscaling a fractionally counted IF5 to this larger set might thus lead to even further improved results.

Using either the percentiles (*PR100*) or the six percentile rank classes used by the NSF (*PR6*) did not add to the considerations in Table 7. Using *PR100*, the reduction is largest in model N5—(130.1 – 55.4)/130.1 = 57.4%—that is, for a two-year citation-window. This is followed by much lower values for TC-FC (11.7%) and TC-FC5 (8.7%). As said, the various models indicate that these variables do not reduce the between-field variance statistically significantly.

d. *Percentile ranks*

|       |     | PR100 |      |      |      | PR6  |      |      |      |
|-------|-----|-------|------|------|------|------|------|------|------|
|       |     | IC2   | FC2  | IC5  | FC5  | IC2  | FC2  | IC5  | FC5  |
| PR100 | IC2 |       | .978 | .986 | .956 | .925 | .904 | .915 | .886 |
|       | FC2 | .978  |      | .969 | .983 | .905 | .925 | .899 | .912 |
|       | IC5 | .986  | .969 |      | .973 | .914 | .898 | .925 | .899 |
|       | FC5 | .956  | .983 | .973 |      | .887 | .913 | .901 | .925 |
| PR6   | IC2 | .868  | .851 | .859 | .837 |      | .924 | .952 | .894 |
|       | FC2 | .851  | .868 | .846 | .858 | .943 |      | .914 | .941 |
|       | IC5 | .860  | .847 | .868 | .849 | .965 | .938 |      | .914 |
|       | FC5 | .837  | .858 | .847 | .868 | .924 | .961 | .939 |      |

**Table 8:** Spearman rank-order correlations (upper triangle) and Pearson correlations (lower triangle) among percentile rankings; $N = 3,705$; all correlations are statistically significant at the level $p < 0.001$.

Table 8 shows the Spearman rank-order correlations (upper triangle) and Pearson correlations (lower triangle) for a selection of integer and fractionally counted percentile rank classes, first in the scheme of *PR100* and then also in terms of *PR6*. The table teaches us some things about the differences between the two evaluation schemes which accord with analytical expectations, but nevertheless deserve to be mentioned.

First, in the upper-left quadrant, there is no difference between Pearson and Spearman correlations in the case of percentiles. Of course, the transformation into percentiles does not affect the ordering. However, values of the Pearson correlation coefficients in the bottom-left (off-diagonal) compartment are systematically lower than the corresponding Spearman rank-order correlations in the upper-left quadrant. The Pearson correlation is sensitive to the breaks introduced by the non-linear binning into six categories, whereas this matters less for rank-order correlations.

For example, the difference between a quantile of 75.01 versus 74.99 is translated into a difference between 3 and 2 on a scale of six when using *PR6*. This loss in precision is the trade-off for the reduction of complexity in the (normative) evaluation scheme: a score of 99.01 then belongs to the highest excellence class (top-1%), whereas 98.99 does not, independently of the error term in the measurement (Leydesdorff, 2012c; cf. Schreiber, 2012b).

|    | IC                  | FC                  | FC5                 | FC2                 |
|----|---------------------|---------------------|---------------------|---------------------|
| 1  | Anal Chem           | Adv Mater           | Adv Mater           | Adv Mater           |
| 2  | Angew Chem Int Edit | Anal Chem           | Anal Chem           | Anal Chem           |
| 3  | Appl Phys Lett      | Angew Chem Int Edit | Angew Chem Int Edit | Angew Chem Int Edit |
| 4  | Astron Astrophys    | Appl Phys Lett      | Appl Phys Lett      | Appl Phys Lett      |
| 5  | Biochemistry-Us     | Biochemistry-Us     | Astron Astrophys    | Astron Astrophys    |
| 6  | Blood               | Blood               | Blood               | Astrophys J         |
| 7  | Cancer Res          | Brit Med J          | Cancer Res          | Blood               |
| 8  | Cell                | Cancer Res          | Cell                | Cell                |
| 9  | Chem Rev            | Cell                | Chem Commun         | Chem Commun         |
| 10 | Circulation         | Chem Rev            | Circulation         | Circulation         |
| 11 | Inorg Chem          | Circulation         | Environ Sci Technol | Inorg Chem          |
| 12 | J Am Chem Soc       | Inorg Chem          | Inorg Chem          | J Alloy Compd       |
| 13 | J Appl Phys         | J Am Chem Soc       | J Am Chem Soc       | J Am Chem Soc       |
| 14 | J Biol Chem         | J Appl Phys         | J Appl Phys         | J Appl Phys         |
| 15 | J Chem Phys         | J Biol Chem         | J Biol Chem         | J Biol Chem         |
| 16 | J Clin Invest | J Chem Phys | J Chem Phys | J Chem Phys |
| 17 | J Clin Oncol | J Clin Invest | J Clin Oncol | J Clin Oncol |
| 18 | J Geophys Res | J Clin Oncol | J Geophys Res | J Geophys Res |
| 19 | J Immunol | J Geophys Res | J Immunol | J Immunol |
| 20 | J Neurosci | J Immunol | J Neurosci | J Neurosci |
| 21 | J Org Chem | J Neurosci | J Phys Chem B | J Phys Chem A |
| 22 | J Phys Chem B | J Org Chem | J Phys Chem C | J Phys Chem B |
| 23 | J Virol | J Phys Chem B | Jama-J Am Med Assoc | J Phys Chem C |
| 24 | Jama-J Am Med Assoc | Jama-J Am Med Assoc | Lancet | Lancet |
| 25 | Lancet | Lancet | Langmuir | Langmuir |
| 26 | Langmuir | Langmuir | Macromolecules | Macromolecules |
| 27 | Macromolecules | Macromolecules | Nano Lett | Nano Lett |
| 28 | Nature | Nature | Nature | Nature |
| 29 | New Engl J Med | New Engl J Med | New Engl J Med | New Engl J Med |
| 30 | Nucleic Acids Res | Nucleic Acids Res | Opt Express | Opt Express |
| 31 | P Natl Acad Sci Usa | P Natl Acad Sci Usa | Org Lett | Org Lett |
| 32 | Phys Rev A | Phys Rev A | P Natl Acad Sci Usa | P Natl Acad Sci Usa |
| 33 | Phys Rev B | Phys Rev B | Phys Rev A | Phys Rev A |
| 34 | Phys Rev D | Phys Rev D | Phys Rev B | Phys Rev B |
| 35 | Phys Rev Lett | Phys Rev Lett | Phys Rev D | Phys Rev D |
| 36 | Science | Science | Phys Rev Lett | Phys Rev Lett |
| 37 | Tetrahedron Lett | Tetrahedron Lett | Science | Science |

**Table 9**: The 37 top-1% journals of the 3,705 journals under study using different indicators.

Table 9 shows the top-1% (37 of the 3,705 journals) using different schemes. Given the large set, the use of fractional or integer counting may make for marginal differences in terms of percentage classes. In the case of *PR6,* such differences can have more impact than in the case of *PR100*.

In sum, if one wishes clear sets for policy purposes, then a scheme of six percentile rank classes meets the objective. For a more nuanced decision, for example, in the case of library management, one may wish to use percentiles or, even more easily, the total cites of journals without these transformations (Bensman, 1996, 2007). The ranks based on numerators are highly correlated; however, a very different dimension (IFs or c/p) is indicated after division by the number of citable items (Leydesdorff, 2009).

Given that we are studying here the entire journal set, no reference set is defined against which to test whether these percentile values are above or below expectation. All averages of *PR100* are therefore between 49.9 and 50.0 (and all medians, too). All averages of *PR6* are between 1.909 and 1.910 (Bornmann & Mutz, 2011). Percentiles and percentile rank classes are used here only to normalize the skewed citation distributions. The slight deviance from the expected values of 50 and 1.91, respectively, are caused by our handling of tied ranks as discussed above (cf. Schreiber, 2012, in press; Leydesdorff, 2012a, in press). By using the Integrated Impact Indicator (I3) at the article level, however, one would obtain for each journal a set of values which can be tested against the journals in the class (Leydesdorff & Bornmann, 2011b; Leydesdorff, 2012c, in press). However, journals were not further decomposed at the document level in this study.

**Conclusions**

We raised two questions with respect to the across-field comparison of the citation impacts of journals using the set of 3,705 journals contained on the CD-ROM of the SCI: (*i*) does fractional counting matter, and if so, for which time horizons, and (*ii*) how is the rating affected by transforming the skewed distributions into percentile values as a non-parametric scale?

1. Fractional counting of the citations diminishes the between-field level differences by more than 90% if the five-year citation window is used. This is considerably more than the approximately 80% reduction that we found both in this study and the previous one (Leydesdorff & Bornmann, 2011a) when using IF2.

2. The fractionation is best pursued with reference to the number of references in the shell of the past five years. Extension to the full set—that is, including older years—worsens the intended normalization because the longer-term tales of the distributions introduce field-specificity.

3. The transition to percentile rank classes did not add to the evaluation of journals in terms of times cited. The rankings are relatively robust against these transformations. One risks adding artificial breaks between classes (and thus error) when introducing a smaller number of classes more than when using the full distribution.

In other words, these transformations of the data seem unnecessary from the perspective of the objective of reducing between-field differences. Although the orders may change, it is unclear whether this leads to an improvement because all resulting indicators lead to statistically significant differences when compared across the eleven broad fields.

We did not test whether there is an optimal citation window other than two or five years. One could consider the citation window as a continuous variable (discretized into publication years) and perhaps find an optimum (cf. Rousseau, 2006). However, the optimum may also change over time (Althouse *et al.*, 2009). It seems to us that we have specified in this study the design for such a further (perhaps more industrial) optimization.

**Discussion and further perspectives**

Our conclusion to use the fractionally counted IF5 as the best currently available alternative remains a statistical answer to the problem of choosing an indicator for the comparison across fields. The validity of this answer depends on the research question. In other settings, one may wish to use *PR100* or *PR6*, for example, for the evaluation of institutional units of analysis (Leydesdorff et al., 2011). One may also wish to validate the resulting rankings against faculty ratings (Bensman, 1996) or in relation to the results of peer review (Bornmann, 2011).

Fractional counting in these cases first counteracts between-group differences, for example, in the case of evaluating multi- or interdisciplinary units such as universities and institutions (Leydesdorff & Shin, 2011; Rafols et al., 2011; Zhou & Leydesdorff, 2011; cf. Radicchi & Castellano, 2012; Sirtes, 2012). Secondly, fractional counting in terms of the citing articles as the relevant audience sets frees the analyst from the need to define the fields in terms of journals or other categorical schemes by making this question empirical. As Glänzel *et al*. (2011) noted, this methodology for the classification in terms of relevant (citing) audiences leaves the problem of classifying uncited papers unsolved.

While our research was in progress, Radicchi & Castellano (2012) published a proposal for an alternative to using fractional counting, by elaborating on the relative citation indicator that Radicchi, Fortunato, & Castellano (2007) had proposed as a rescaling of the distribution on a universal curve. These authors claim that the citation values in a group (e.g., a field) have first to be rescaled by dividing each citation count by the average of the citation values in the group. In formula format: $c_f = c/c_0$, in which $c$ is the number of citations for each case and $c_0$ the average of this variable in the group to which the case belongs. After this normalization, citation distributions would be "universally" comparable across fields. In the meantime, Waltman *et al*. (2012) questioned this claim of "universality." We would have liked to include this other normalization into our comparison (cf. Sirtes, 2012), but our design of nonparametric variance-components analysis was not sufficiently comparable with the proposed normalization (to the arithmetic mean), and we therefore decided to postpone this further evaluation to a future study.

**Acknowledgement**
We acknowledge Thomson-Reuters for access to the data. We thank Kim Hamilton for providing us with the journal list of PatentBoard, and NSF for giving permission.